\documentstyle[preprint,aps]{revtex}

\tightenlines

\begin{document}

\draft

\title{Supersymmetric Harmonic Maps into Lie Groups}

\author{Fergus O'Dea}

\address{Physics Department, The University of Texas at Austin,
Austin, Texas 78712}

\date{\today}

\maketitle

\begin{abstract}

We look at the supersymmetric generalization of harmonic maps
into Lie groups, known to physicists as the chiral model.
Explicit solutions to the equations are found
and examined using Backlund transformations.

\end{abstract}

\section{Introduction}

\subsection{Motivation}

Harmonic maps into Lie groups
have been of interest to physicists since the
1970's under the name of chiral models (we use the terms interchangeably)
and studied them as toy models for gauge theories. The concept of
a harmonic map generalises the concept of a geodesic. 

The
harmonic maps in this paper are from a simply connected
2-dimensional Riemannian domain to a Lie group and are therefore
two dimensional analogues of geodesics.

Harmonic maps in two dimensions are special because they have
the property of being an "integrable system". This term is used
loosely in the literature, and no precise definition exists.
Various fundamental properties of an integrable system are
recognised, but normally it is a differential equation, with a
large symmetry group, which is solvable by algebraic means.  We
will assume for our purposes the existence of a Lax pair to be
sufficient.

The chiral model is an example of a relativistically invariant
integrable non-linear p.d.e., and as such is important to
physicists as a physical model. It is related to the chiral
symmetries which feature in particle physics, such as the study
of strong interactions at low energies.

Most of the known integrable theories are two dimensional. The
classic examples are the Korteweg-de Vries and Sine-Gordon
models, and the integrability problem for two-dimensional field
theories has been studied systematically using Lax equations and the theory of
affine Lie algebras.

Harmonic maps in two dimensions have been 
studied in the past ten years using a
paramatrised Lax equation. \cite{U}.

In our analysis, we draw heavily on the uniton solutions
constructed using a parametrised Lax pair by Uhlenbeck in 1989.
She showed that harmonic maps into a unitary group can be
factorised into a product of these simple maps.  Harmonic maps
into unitary groups can therefore be characterised by a certain
number, the number of these factors, known as the uniton number.

The chiral model has a
relationship to several different non-linear field theories, such
as the non-linear $\sigma$ model. They are members of a class of
models constructed on special coset spaces or homogeneous spaces, the symmetric
spaces.

Supersymmetric field theories are the most promising candidates
to extend the standard model of strong and electroweak
interactions and were first studied in the context of string
theory. 
The question then arises
as to what are the analagous results for the supersymmetric
chiral model.

\subsection{Outline}

The outline of this paper is as follows: In section II, we define
harmonic maps, in setions III and IV we describe the connections
between harmonic maps and other integrable systems, and in
sections V through VII we study supersymmetric harmonic maps.

In section II, we give a very brief introduction to the theory of
harmonic maps. We look at the particular case of harmonic maps
into Lie groups, and review fundamental results of Uhlenbeck,
including the uniton solutions. 

In section III, we begin our development of a theory of
supersymmetric harmonic maps or superharmonic maps with a brief
review of the formalism of N=1 supergeometry. We present our
definition of the supercurvature and we prove that the definition
permits the construction of unique super Lax pair solutions and
extended solutions. In section IV, we find the supersymmetric
analogue of Uhlenbeck's uniton solutions and write down explicit
formulae for these special simple superharmonic maps into unitary
supergroups. We conclude in section V by developing Backlund
transformations between these maps as a tool to help analyse
them. In doing so, we construct a representation on extended
superharmonic maps and use this write down some example Backlund
transformations for the superuniton case, highlighting some
unique features.

\section{Harmonic Maps}

The most common definition of the harmonic map is as the critical
point of the energy functional, the
integral of the energy density over the domain.  

There are some simple examples of harmonic maps.  If the target manifold is
the space of real numbers, then $f$ is just a harmonic function on $M$, i.e. it vanishes under
the Laplacian operator.  If the domain is an interval on the Real line, the equation
 becomes that of a geodesic.

After elaborating on these definitions and looking at a few examples, we consider in detail
the case of harmonic maps
into Lie groups or the chiral model. In later sections, we will describe the
supersymmetric version.

\subsection{The energy of a harmonic map}

The energy density of a map $f:M\to N$ between two 
Riemannian manifolds with metrics $g$ and $h$ respectively is defined as
half the trace of the pull-back $f^*h$ with respect to the metric $g$ of $M$:

\begin{equation}
$$e(f)={1 \over 2}Tr_g(f^*h)$$
\end{equation}
where $\left\{ {e_i} \right\},i=1,...,\dim M$.    The integral of the
energy density over $M$ yields the energy of the map:

\begin{equation}
$$E(f)={1 \over 2}\int\limits_M {\sqrt g Tr_g(f^*h)}$$
\label{energy}
\end{equation}
When $M$ is compact, the energy is finite.

{\bf Definition}  {\it A map $f:M\to N$ is harmonic if and only if it is an extremal
of the energy integral $E$}

\subsection{Harmonic maps into Lie groups}

At this point we look at maps $f:M \to G$, where $G$ is the real form
of a complex Lie group and $M$ is a Riemann surface. The Maurer-Cartan form $\theta$ 
 maps left-invariant vector fields to their corresponding Lie algebra element,
and because of identities relating the structure constants of the group,  
satisfies the Maurer-Cartan structure equation:

\begin{equation}
$$d\theta +{1 \over 2}[\theta \wedge \theta ]=0$$
\end{equation}

The pull-back $B =f^*\theta $ is an $L(G)$ valued 1-form on $M$ satisfying 

\begin{equation}
$$dB +{1 \over 2}[B \wedge B ]=0$$
\end{equation}
If we view this as the equation for a flat connection on $M$, we can write
$A =f^{-1}df$.
Write the energy of a map $f:M\to G$, using (\ref{energy}) as

\begin{equation}
$$E(f)={1 \over 2}\int\limits_M {\sqrt g\sum\limits_i {\left| {f^{-1}{{\partial f} \over {\partial \xi ^i}}} \right|}}^2$$
\end{equation}

If we find the Euler-Lagrange equations for this integral, or use the
result \cite{Pl} that a map $f$ into a Lie group is
harmonic if and only if 
$d^*(f^*\theta )=0$ , we arrive at
the harmonic map equations in this case as:

\begin{equation}
$${\partial  \over {\partial x}}\left( {f^{-1}{{\partial f} \over {\partial x}}} \right)+{\partial  \over {\partial y}}\left( {f^{-1}{{\partial f} \over {\partial y}}} \right)=0$$
\end{equation}

Let $2A=B$. The above equation can be written in characteristic co-ordinates as:

\begin{equation}
$$\partial_{\bar z} A +\partial_z {\bar A} =0$$
\label{harmalpha}
\end{equation}

By the definition of $A$ we have

\begin{equation}
$$\partial_{\bar z} A_z -\partial_z A_{\bar z} +2\left[ {A_{\bar z} ,A_z } \right]=0 \ .$$
\label{defalpha}
\end{equation}
The above two equations (\ref{harmalpha}) and (\ref{defalpha})
are equivalent to a single equation (and its dual):

\begin{equation}
$$\partial_{\bar z} A_z +\left[ {A_{\bar z} ,A_z } \right]=0$$
\end{equation}

We need the following theorem from \cite{U} in all
that follows:

{\bf Theorem 2.1}  	{\it Let $\Omega$ be simply connected and $A:\Omega
\rightarrow\ T^*(\Omega)\otimes{\bf g}$. Let $2A=f^{-1} df$. Then f is
harmonic if and only if the curvature of the connection in $\Omega \times{\bf C}^N$}
\begin{equation}
D_{\lambda}=\left( {\partial_{\bar z}+(1-\lambda ){\bar A},\partial _z+(1-\lambda
^{-1})A_z} \right)
\label{eq3.1}
\end{equation}
 
{\it vanishes for all }$\lambda\in{\bf C}^*={\bf C}-0$

{\bf Proof:}

This is proved by writing out the curvature equations and expanding in $\lambda$.

The problem is also described as a pair of simultaneous linear equations
which trivialise the connection:

\begin{equation} 
$$\partial _{\bar z }g_\lambda =(1-\lambda ) g_\lambda A_{\bar z } ,\ \ \ \ \ \partial _z g_\lambda =(1-\lambda ^{-1})g_\lambda A_z  $$
\label{ddef}
\end{equation} 

The $g$ which satisfy these equations can be
uniquely chosen to satisfy  the properties
described in the following theorem \cite{U}:

{\bf Theorem 2.2}  {\it Let $f$ be harmonic and $f_0(p)\equiv I$ for some $p \in \Omega$.
Then there exists a unique $g_\lambda :C\times \Omega \to {\bf G}$ satisfying (\ref{ddef}) 
with}

\begin{equation}
$$\matrix{(a)\ g_1\equiv I\hfill\cr
  (b)\ g_{-1}=f\hfill\cr
  (c)\ g_\lambda (p)=I\hfill\cr
  \hfill\cr}$$
\label{thmg}
\end{equation}

{\it Furthermore, if $f$ is unitary, $g_\lambda $ is unitary for $\left| \lambda  \right|=1$.}

{\bf Theorem} The converse is also true: if $g_\lambda :C\times \Omega \to G$ is analytic and 
holomorphic in the first variable, $g_1 \equiv I$
and the expressions
\begin{equation}
$${{\partial _{\bar \theta }g_\lambda \cdot \left(
{g_\lambda } \right)^{-1}} \over {1-\lambda
}}\matrix{{}\cr {}\cr }\matrix{{}\cr
{}\cr
}{{\partial _\theta g_\lambda \cdot \left(
{g_\lambda }
\right)^{-1}} \over {1-\lambda ^{-1}}}$$
\end{equation}
are constant in $\lambda $, then  $f=g_{-1}$ is
harmonic. 

These extended solutions $g_\lambda (x)$can be expanded in a Laurent series in $\lambda$ about either zero or $\infty$.
Harmonic maps can be classified according to their {\bf uniton number}.

{\bf Definition} The {\bf uniton number} is 
the highest power of $\lambda$ in this Laurent series expansion, possibly $\infty$.

The maps we look at are local to $\Omega \subset {\cal C}$.  We assume that
these maps in ${\cal C}$ can be extended over the point at $\infty$ to maps into $S^2 = {\cal C} \cup \infty$
Then these equations have finite-energy solutions.

{\bf Definition}  An {\bf n-uniton} is a harmonic map {\it $f:\Omega \to U\left(N\right)$}
which has an extended solution

\begin{equation}
$$g_\lambda :C^*\times \Omega \to GL(N)$$
\end{equation}

with

\begin{eqnarray}
&(a)&\ g_\lambda =\sum\limits_{\alpha =0}^n
{T_\alpha \lambda ^\alpha \ for\ T_\alpha }:\Omega
\to L(G)\hfill \\
  &(b)&\ g_1=1\hfill \\
  &(c)&\ g_{-1}=qf^{-1}\ for\ q\in U(N)\
constant \\
  &(d)&\ \left( {g_{\bar \lambda }}
\right)^*=\left( {g_{\lambda ^{-1}}} \right)^{-1}
\end{eqnarray}

For $n=0, g_\lambda \equiv I$ is the only extended
solution, which represents 
$f \equiv q^{-1}$ or the constant harmonic maps.

We will also take note of the fact, also shown
in \cite{U}, that all these harmonic 
 maps are defined into totally geodesic Grassmanian submanifolds.
We must consider maps into

\begin{equation}
$$G_{k,N}\subset G_{}=U(N)$$
\label{}
\end{equation}

given by

\begin{equation}
$$G_{k,N}=\left\{ {\phi \in U\left( {N}
\right):\phi ^2=I\ with\ k-dimensional\ +1\
eigenspace} \right\}$$
\label{incdef}
\end{equation}

We identify $\phi$ with the $k$-dimensional subspace corresponding to the $+1$ eigenspace.
Since the embedding is geodesic, we can identify maps $s$ into $U(N)$ 
satisfying $s^2=1$ as harmonic because they represent hermitian projections $\pi $
on a $k$-dimensional sub-bundle in $\Omega \times {\bf C}^N$.  These harmonic maps 
are given by 

\begin{equation}
$$s=\pi -\pi ^\bot=2\pi -I$$
\label{}
\end{equation}

 So ends our accelerated
review of the theory of harmonic maps, at least
insofar as it applies in this work.

\section{Superharmonic maps into super Lie groups}

In this section we make a study of the chiral model in N=1
superspace.
 After reviewing some elementary principles
of supergeometry, we construct a superspace version of the Lax pair construction.
This allows us to examine integrable systems in superspace, and in particular
we define a superharmonic map into a (super) Lie group.
Extended solutions to the associated linear problem are defined, and this will lead us into
a study of the simplest superharmonic maps and their solutions in the next
section.

\subsection{Supergeometry}
 
We work throughout in complexified superspace with co-ordinates
$(z,\bar z;\theta ,\bar \theta )$.  $(\theta ,\bar \theta )$
are anticommuting variables.  Thus, a sign change occurs
when moving such quantities past each other.  
For convenience, we will denote by a hat ${\hat X}$
the operation of flipping the sign  
of the anti-commuting or 'odd' part of any quantity $X$.

The derivates in the anticommuting variables are
 themselves anticommuting quantities, as are the superderivates

\begin{equation}
$$D_\theta =\partial _\theta +\theta \partial _z,\ \ \ D_{\bar \theta }=\partial _{\bar \theta }+\bar \theta \partial _{\bar z}$$
\label{}
\end{equation}

The superderivates are related to the regular
space derivatives in a simple and easily verifiable manner:
\begin{equation}
$$D_\theta ^2=\partial _\theta ,\ \ \ \ \ \ D_{\bar \theta }^2=\partial _{\bar \theta }$$
\label{}
\end{equation}

Any quantity defined over superspace can be expanded in a Taylor series over the anti-commuting co-ordinates

\begin{equation}
$$X=x_0+x_+\theta +x_-\bar \theta +x_2\theta \bar \theta $$
\label{comps}
\end{equation}

and it is convenient to denote this in vector
notation as

\begin{equation}
$$X=\left( {\matrix{{x_0}\cr
{x_+}\cr
{x_-}\cr
{x_2}\cr
}} \right)$$
\label{vectornot}
\end{equation}

\subsection{Lax pair for superspace}

A unique feature of our analysis is the definition
of the {\bf supercurvature} for a connection $(A_\theta ,A_{\bar \theta })$ defined over superspace,
denoted $F(A_\theta ,A_{\bar \theta })$.  

It is distinguished from the simple supercommutator
by minor but essential sign changes, which are recorded wuith the hat notation
we described above.

{\bf Definition 5.1}  The {\bf supercurvature} of 
the connection $(A_\theta ,A_{\bar \theta })$ is defined as

\begin{equation}
$$F(A_\theta ,A_{\bar \theta })=D_{\bar \theta }A_\theta +D_\theta A_{\bar \theta }+\hat A_{\bar \theta }A_\theta +\hat A_\theta A_{\bar \theta }$$
\label{curv}
\end{equation}

This is also the definition we use when we refer to Lax pairs in superspace.
Any analysis now requires that such Lax pairs make sense in the usual
way of integrable systems, that they are equivalent to an associated linear problem.
Hence the following theorem.

{\bf Theorem 5.1}  {\it The vanishing of the supercurvature $F(A_\theta ,A_{\bar \theta })$}
{\it is a necessary and sufficient condition for the existence and uniqueness
of a trivialization of the connection $(A_\theta ,A_{\bar \theta })$ in connected superspace, that is to say, a solution to the
equations}

\begin{equation}
$$D_\theta G=A_\theta G,\ \ \ \ \ D_{\bar \theta }G=A_{\bar \theta }G$$
\label{dg=ag}
\end{equation}

{\it exists if and only if our connection is flat according to our definition, where
$G$ takes its values in some Lie group or supergroup and $A$ in the
corresponding algebra.}

{\bf Proof}
Clearly, since the superderivative anticommutes, we have
\begin{equation}
$$D_\theta D_{\bar \theta }G+D_{\bar \theta }D_\theta G=0$$
\label{dd+dd}
\end{equation}

so that, by (\ref{dg=ag}), we get
\begin{equation}
$$D_\theta (A_{\bar \theta }G)+D_{\bar \theta }(A_\theta G)=0$$
\label{dag}
\end{equation}

Expanding this, and bearing in mind that moving the superderivative
past a term changes the sign of the odd components, we find that the 
expression 
(\ref{curv}) vanishes.

It is more laborious to prove the theorem in the other direction,
to show that flat superconnections can be trivialised but it
can be shown that the problem reduces to a standard one
for a trivial connection in the body component alone.
By the existence theorem for PDE, a unique solution exists for $g_0$ and hence $G$ once we prescribe its
value at some base point, provided that the 
compatibility conditions for the second and seventh of (\ref{8set}) are satisfied,
that is, that the curvature of this connection vanishes in the traditional sense.

It is clear on examination of these calculations
that our definition of the Lax pair  (the
supercurvature) is fully determined
 and is not merely a
convenient choice.  With this understood we press on to define a superharmonic map
from superspeace into a super Lie group.

\subsection{Superharmonic maps}
Firstly, we will need a couple of identities which are easily checked.  Again, the hat indicates 
that the odd components of the term are given opposite sign.

\begin{equation} 
$$\matrix{D_\theta M^{-1}=-\hat M^{-1}\cdot D_\theta M\cdot M^{-1}\hfill\cr
  D_\theta (MN)=(D_\theta M)N+\hat M(D_\theta N)\hfill\cr}$$
\label{idents}
\end{equation}

{\bf Definition} A {\bf superharmonic map} $S:\Omega \to G$ 
from $N=1$ superspace into a (super) Lie group $G$
 is a solution of
the following equation:

\begin{equation} 
$$D_\theta \left( {D_{\bar \theta }S\cdot S^{-1}} \right)-D_{\bar \theta }\left( {D_\theta S\cdot S^{-1}} \right)=0$$
\label{shmdef}
\end{equation}

Clearly, all harmonic maps are trivially superharmonic.
A supersymmetric analysis of superharmonic maps
analogous to the material presented in section II now follows.
As in the non-supersymmetric case, we can introduce a spectral 
parameter to encode this definition of a superharmonic map 
in terms of a single Lax pair expression. We will need this result, 
expressed in the following theorem, in all that follows:

{\bf Theorem 5.2}  {\it Let $\Omega$ be simply
connected and $A:\Omega
\rightarrow\ T^*(\Omega)\otimes L(G)$. 

Then $2A_\theta ,_{\bar \theta }=D_\theta ,_{\bar \theta }S\cdot S^{-1}$, 
where $S$ is superharmonic if and only if the supercurvature}
\begin{equation}
F\left( {(1-\lambda ^{-1})A_\theta ,(1-\lambda
)A_{\bar \theta }} \right)=0
\label{thm}
\end{equation}
 
{\it vanishes for all} $\lambda\in{\cal C}^*={\cal C}-0$

In terms of coefficients of $\lambda$,  we have

\begin{equation} 
$$\matrix{\lambda \left( {D_\theta A_{\bar \theta }+\hat A_{\bar \theta }A_\theta +\hat A_\theta A_{\bar \theta }} \right)+\lambda ^0\left( {D_\theta A_{\bar \theta }+D_{\bar \theta }A_\theta +2(\hat A_{\bar \theta }A_\theta +\hat A_\theta A_{\bar \theta })} \right) \hfill\cr
  +\lambda ^{-1}\left( {D_{\bar \theta }A_\theta +\hat A_{\bar \theta }A_\theta +\hat A_\theta A_{\bar \theta }} \right)=0\hfill\cr}$$
\end{equation}

{\bf Proof} It is a simple matter to show that this is equivalent to the
supersymmetric form of the chiral equation and an identity arising from the
definition of $A_\theta$, $A_{{\bar \theta }}$.

By theorem 1 and since $\Omega$ is simply connected,
 we can explore solutions to these equations by solving the trivialisation problem.
These are solutions $G_\lambda $ to the
simultaneous equations

\begin{equation} 
$$D_{\bar \theta }G_\lambda =(1-\lambda )A_{\bar
\theta }G_\lambda ,\ \ \ \ \ D_\theta G_\lambda
=(1-\lambda ^{-1})A_\theta G_\lambda $$
\label{Gdef}
\end{equation}

which are valid for all values of the spectral parameter $\lambda$, of which $G^\lambda$ is a function.
  
Solutions are prescribed uniquely by setting the body component,
$\left( {g_\lambda } \right)_0\left( p \right)$
at any base point
$p\in
\Omega
$.    We can safely normalise our solution so that
 $s_0(p)=I$ at $p$.  We can do this 
since $\tilde S=S\cdot s_0^{-1}(p)$ is superharmonic if $S$ is, and $\tilde s_0(p)=I$.
We then choose $g_0^\lambda (p)=I$$\left(
{g_\lambda } \right)_0\left( p \right) =I$.

Also, if $\lambda =1$, $D_\theta G^1=D_{\bar \theta }G^1=0$, for which a Taylor
expansion in $\theta, {\bar \theta}$ reveals that $dg_0^1=0$, all other components vanishing identically.
So $G^1\equiv I$.  

At this point we can write down a theorem:

{\bf Theorem 5.3}  {\it If $S$ is superharmonic and
$s_0(p)\equiv I$, then there exists a unique
$G^\lambda :C\times \Omega \to {\bf G}$ satisfying
(\ref{Gdef})  with}

\begin{equation}
$$\matrix{(a)\ G^1\equiv I\hfill\cr
  (b)\ G^{-1}=S\hfill\cr
  (c)\ G^\lambda (p)=I\hfill\cr
  \hfill\cr}$$
\label{thmG}
\end{equation}

{\it Also if $S$ is unitary and a $c$-number
(commuting quantity),
$G^\lambda$ is unitary for $\left| \lambda 
\right|=1$.}

{\bf Proof}  It only remains to check the final statement, which follows from the fact that

\begin{equation} 
$$-D_{\bar \theta }(G^\lambda )^{-1}=(1-\lambda )(\hat G^\lambda )^{-1}A_{\bar \theta },\ \ \ \ \ \ -D_\theta (G^\lambda )^{-1}=(1-\lambda ^{-1})(\hat G^\lambda )^{-1}A_\theta $$
\label{Gdefinv}
\end{equation}

As in section II, the converse is also true.  That is, if $G^1=I$
and  the quantities

\begin{equation}
$${{D_{\bar \theta }G^\lambda \cdot \left( {G^\lambda } \right)^{-1}} \over {1-\lambda }},\ \ \ \ {{D_\theta G^\lambda \cdot \left( {G^\lambda } \right)^{-1}} \over {1-\lambda ^{-1}}}$$
\label{}
\end{equation}

are constant in $\lambda $, then  $S=G^{-1}$ is superharmonic.

The point of all this is that we can now study superharmonic maps using
only the data encoded by the $G^\lambda$.  
 In particular, we can expand
the $G^\lambda$ in a Laurent series as a function of $\lambda$

\begin{equation}
$$G^\lambda =\sum\limits_{\alpha =-\infty }^\infty  {T_\alpha }\lambda ^\alpha $$
\label{laurent}
\end{equation}

where $T_\alpha$ are Lie algebra-valued functions on superspace.  

We use these tools to find explicit solutions to a simple class of superharmonic maps
in the following section.

\section{Superharmonic map solutions in the unitary case}

We seek to understand the relationships of superharmonic maps into Lie groups
to  the traditional harmonic maps and other integrable systems.
For the case of the the 
simplest finite-action superharmonic maps
 into $SU(M/N)$, which we call superunitons,
we find expressions for their solution in terms of 
a holomorphic projector and a related quantity.

By analogy with the non-supersymmetric case, any
finite-energy  superharmonic map can be factorised
in terms of these maps, and for $SU(1/1)$ it is
known that these are the only such maps.

We first look at the construction of maps into superspace
Grassmannians, which will give us  the mathematical equipment
we need for the rest of this analysis.

\subsection{Superunitons}

As in the non-supersymmetric theory, 
our study of superharmonic maps is made by examining maps 
into Grassmannians.  In this case, we must be conscious of the
presence of the superspace variables by looking at maps from superspace
into supersymmetric Grassmannians, the space of $k/l$-planes in 
$M/N$ superspace.

Then consider the inclusion

\begin{equation}
$$G_{k,l;M,N}={{U\left( {M/N} \right)} \over {U(k/l)\times U(M-k/N-l)}}\subset U\left( {M/N} \right)$$
\label{inc}
\end{equation}

defined by 

\begin{equation}
$$G_{k,l;M,N}=\left\{ {\phi \in U\left( {M/N}
\right):\phi ^2=I\ with\ k/l\ dimensional\ +1\
eigenspace} \right\}$$
\label{incdef}
\end{equation}

The embedding $G_{k,l;M,N}\subset U\left( {M/N} \right)$ is totally geodesic, and we have

{\bf Proposition 6.1}  {\it The combined map} $\phi
S: \Omega \buildrel S \over \longrightarrow
G_{k,l;M,N}\buildrel \phi  \over \longrightarrow
U\left( {M/N} \right)$ {\it is superharmonic if
and only if $S$ is superharmonic.}

So we study superharmonic maps

$S:\Omega \to U\left( {M/N} \right)$ for which
$S=S^{-1}$. The maps are given by

\begin{equation}
$$S=\left( {\Pi -\Pi ^\bot } \right)=\left(
{2\Pi -1} \right)$$
\end{equation}

where $\Pi$ is the (super)Hermitian projection of
rank
$k,l$ on a $k/l$-dimensional sub-bundle at each
$q\in \Omega$.  

Algebraically we can identify $\Pi$ by noting that

\begin{eqnarray}
&(a)&\ \Pi ^*(q)=\Pi (q)\ \ for\ all\ q\in
\Omega \\
  &(b)&\ \Pi ^2(q)=\Pi (q)\ \ for\ all\ q\in
\Omega \\
  &(c)& \Pi (q)\ \ has\ rank\ k/l\ at\
every\ point\ q\in \Omega
\label{}
\end{eqnarray}

 We also note that $\Pi ^\bot = \left(1-\Pi \right)$
is the Hermitian projection on the orthogonal bundle.

We now define a discrete series of elemental superharmonic maps, which we call superunitons,
as a supersymmetric extension of the constructions of \cite{U}.
The level of complexity of these maps as well as the energies are defined 
by the minimum number of terms needed in the
expansion an extended solution.

{\bf Definition}  An {\bf n-superuniton} is a superharmonic map {\it $S:\Omega \to U\left( {M/N} \right)$}
which has an extended solution

\begin{equation}
$$G^\lambda :C^*\times \Omega \to GL(M/N)$$
\end{equation}

with

\begin{equation}
$$\matrix{(a)\ G^\lambda =\sum\limits_{\alpha =0}^n {T_\alpha \lambda ^\alpha \ for\ T_\alpha }:\Omega \to L(G)\hfill\cr
  (b)\ G^1=1\hfill\cr
  (c)\ G^{-1}=QS^{-1}\ for\ Q\in U(M/N)\ constant\hfill\cr
  (d)\ \left( {G^{\bar \lambda }} \right)^*=\left( {G^{\lambda ^{-1}}} \right)^{-1}\hfill\cr}$$
\end{equation}

For $n=0, G^\lambda \equiv I$ is the omly extended solution, which represents 
$S \equiv Q^{-1}$ or the constant superharmonic maps.

We now examine the 1-superuniton, the simplest
example.

{\bf Proposition 6.2}  {\it $S:\Omega \to U\left(
{M/N} \right)$ is a one-S-uniton if and only if
$S=Q\left( {\Pi -\Pi ^\bot } \right)$ for $Q\in
U(M/N)$, where $\Pi$ is a Hermitian projector and
$\bar D\Pi \cdot \Pi ^\bot =0$.} .

{\bf Proof.}  Clearly, we have $G^\lambda =T_0+\lambda \left( {1-T_0} \right)$

The reality condition gives us

\begin{equation}
$$\left( {I-T_0} \right)^*T_0=0,\ \ \ \ \ \ T_0^*\left( {I-T_0} \right)=0$$
\end{equation}

Combining these equations gives us $T_0^*=T_0$ and $T_0^2=T_0$, so that $T_0=\Pi $.

Apart from the final statement, the theorem is now apparent by inspection.
The superharmonic condition from the end of the previous section
is that expressions for 

\begin{equation}
$${{\bar DG^\lambda \cdot \left( {G^\lambda } \right)^{-1}} \over {\left( {1-\lambda } \right)}},\ \ \ \ \ {{DG^\lambda \cdot \left( {G^\lambda } \right)^{-1}} \over {\left( {1-\lambda ^{-1}} \right)}}$$
\end{equation}

be independent of $\lambda$. The former can be written as

\begin{equation}
$$\matrix{{{\bar D\left( {\Pi +\lambda \left( {1-\Pi } \right)} \right)\cdot \left( {\Pi +\lambda ^{-1}\Pi ^\bot } \right)} \over {\left( {1-\lambda } \right)}}\hfill\cr
  ={{\bar D\left( {\left( {1-\lambda } \right)\left( {\Pi -1} \right)} \right)\cdot \left( {\Pi +\lambda ^{-1}\Pi ^\bot } \right)} \over {\left( {1-\lambda } \right)}}\hfill\cr
  =\bar D\Pi \cdot \Pi +\lambda ^{-1}\bar D\Pi \cdot \Pi ^\bot \hfill\cr}$$
\end{equation}

and the latter as

\begin{equation}
$$\matrix{\ {{-\hat E_\lambda \cdot DE_\lambda ^{-1}} \over {\left( {1-\lambda ^{-1}} \right)}}\hfill\cr
  ={{-(\hat \Pi +\lambda \hat \Pi ^\bot )\cdot D\left( {\Pi +\lambda ^{-1}\left( {1-\Pi } \right)} \right)} \over {\left( {1-\lambda ^{-1}} \right)}}\hfill\cr
  ={{(\hat \Pi +\lambda \hat \Pi ^\bot )\cdot D\left( {\left( {1-\lambda ^{-1}} \right)\left( {1-\Pi } \right)} \right)} \over {\left( {1-\lambda ^{-1}} \right)}}\hfill\cr
  =\Pi \cdot D\Pi +\lambda \hat \Pi ^\bot \cdot D\Pi \hfill\cr}$$
\end{equation}

so our condition for superharmonicity is 

\begin{equation}
$$\bar D\Pi \cdot \Pi ^\bot =\left( {\hat \Pi ^\bot \cdot D\Pi } \right)^*=0$$
\end{equation}

This concludes the proof.  

One-unitons in the classical theory are identified
as simply holomorphic maps due to the fact that
Grassmannian manifolds are Kahler. We find in this
case that some additional data is required to
identify the simplest one-superunitons beyond the
holomorphic body component.  

\subsection{Explicit solutions}

We will again use the vector notation of the previous section
to calculate the superuniton condition in terms of the superspace
components.

We expand the projector $\Pi$ in terms of the supervariables $\left( \Theta, \bar \Theta \right)$ as
$\Pi =\pi _0+\pi _+\theta +\pi _-\bar \theta +\pi _2\theta \bar \theta $ or in vector format as

\begin{equation}
$$\Pi =\left( {\matrix{{\pi _0}\cr
{\pi _+}\cr
{\pi _-}\cr
{\pi _2}\cr
}} \right)$$
\end{equation}

{\bf Theorem 6.3}  {\it The one-S-uniton $S:\Omega
\to U\left( {M/1} \right)$
 is given by a holomorphic projector $p$ into $U(M)$ and an
ancillary
$m$-vector $v$ which are related by the equations}

\begin{equation}
$$v^Tp=v_{,\bar z}^T\left( {1-p} \right)=0$$
\end{equation}

{\bf Proof}

In superspace components, we write the superuniton
condition as

\begin{equation}
$$\bar D\Pi \cdot \Pi ^\bot =\left( {\matrix{{\pi _-}\cr
{-\pi _2}\cr
{\pi _{0,\bar z}}\cr
{-\pi _{+,\bar z}}\cr
}} \right)\cdot \left( {\matrix{{1-\pi _0}\cr
{-\pi _+}\cr
{-\pi _-}\cr
{-\pi _2}\cr
}} \right)$$
\label{}
\end{equation}

\begin{equation}
$$=\left( {\matrix{{\pi _-\left( {1-\pi _0} \right)}\cr
{-\pi _-\pi _+-\pi _2\left( {1-\pi _0} \right)}\cr
{-\pi _-^2+\pi _{0,\bar z}\left( {1-\pi _0} \right)}\cr
{-\pi _-\pi _2+\pi _2\pi _--\pi _{+,\bar z}\left( {1-\pi _0} \right)+\pi _{0,\bar z}\pi _+}\cr
}} \right)=0$$
\label{ad}
\end{equation}

Combining equations \ref{ad}a and \ref{ad}b, we find 

\begin{equation}
$$\pi _{0,\bar z}\left( {1-\pi _0} \right)=\pi _-^2=0$$
\end{equation}

and by duality

\begin{equation}
$$\pi _+^2=0$$
\end{equation}

From this we learn three things: firstly, because
$\pi _{0,\bar z}\left( {1-\pi _0} \right)=0$, we know that $\pi _0$ is a solution of
 the non-supersymmetric one-uniton equations described in section II
-- it is just a holomorphic projector. We also see that $\pi _-$,
$\pi _+$ are nilpotent.

If we now take our group $G$ to be $U(M/1)$, then we can write, without loss of generality, 
and using our knowledge of this group's properties,

\begin{equation}
$$\pi _0=\left( {\matrix{p&0\cr
0&1\cr
}} \right)$$
\end{equation}

\begin{equation}
$$\pi _+=\left( {\matrix{0&0\cr
{v^T}&0\cr
}} \right)$$
\end{equation}

\begin{equation}
$$\pi _-=\pi _+^\dagger =\left( {\matrix{0&{i\bar v}\cr
0&0\cr
}} \right)$$
\end{equation}

where $p$ is a holomorphic projector into $U(M)$, $v$ is an $m$-vector, and $\dagger $ denotes
the super-hermitian adjoint.

Since $\Pi ^2=\Pi $, we have

\begin{equation}
$$\left( {\matrix{{\pi _0^2}\cr
{\pi _0\pi _++\pi _+\pi _0}\cr
{\pi _0\pi _-+\pi _-\pi _0}\cr
{\pi _0\pi _2+\pi _2\pi _0+\pi _+\pi _--\pi _-\pi _+}\cr
}} \right)=\left( {\matrix{{\pi _0}\cr
{\pi _+}\cr
{\pi _-}\cr
{\pi _2}\cr
}} \right)$$
\label{eh}
\end{equation}

so that, combining \ref{eh}c with \ref{ad}a from before, we get

\begin{equation}
$$\pi _0\pi _-=0=\pi _+\pi _0$$
\end{equation}

Explicitly, 
\begin{equation}
$$\left( {\matrix{0&0\cr
{v^T}&0\cr
}} \right)\left( {\matrix{p&0\cr
0&1\cr
}} \right)=\left( {\matrix{0&0\cr
{v^Tp}&0\cr
}} \right)=0$$
\end{equation}

so that $v^\bot =0$

Equation \ref{ad}d now reads, if we write $\pi
_2$ as $\left( {\matrix{a&0\cr 0&d\cr
}} \right)$ and plug in for the other components,

\begin{equation}
$$\left( {\bar D\Pi \cdot \Pi ^\bot }
\right)_2=\left( {\matrix{0&{-i\bar vd+ia\bar
v}\cr {-v_{,\bar z}^T\left( {1-p} \right)}&0\cr }}
\right)=0$$
\end{equation}

so that our super-uniton is determined by the additional parameter $v$ and
the condition

\begin{equation}
$$v^Tp=v_{,\bar z}^T\left( {1-p} \right)=0$$
\end{equation}

This ends the proof.

 \ \

We can actually go further and construct an
explicit solution for $\Pi $ in terms of the
holomorphic  projector $p$ and the new ancillary
data.  The complete super-uniton is written explicitly as

\begin{equation}
$$\Pi =\left( {\matrix{{p-i\bar vv^T\theta \bar \theta }&{i\bar v\bar \theta }\cr
{v^T\theta }&{1-iv^T\bar v\theta \bar \theta }\cr
}} \right)$$
\end{equation}

\section{Backlund transformations for
superharmonic maps}

Knowing extended solutions to the superharmonic map equations,
we can generate others using Backlund transformations of the kind
described in \cite{U}. 

A
Backlund transformation is a method of obtaining new solutions of partial
differential equations from old solutions by solving ordinary differential
equations.   In this case, our transformation is obtained by using 
factorisations of solutions extended over
maps on overlapping regions of the 2-sphere.  We will see that the relationship between
these factorisations allows us to transform between
solutions 'of simplest type'.

\subsection{A representation on holomorphic maps}

The factorization we want to use is a modification of the standard Birkhoff 
factorization.  In this case, the 2-sphere $C^*\cup {0}\cup \left\{ \infty  \right\}$
is divided into two overlapping regions:

\begin{equation}
$$S_\varepsilon ^+=\left\{ {\lambda :\left| \lambda  \right|\ge \left( {1+\varepsilon } \right)^{-1}} \right\},\ \ \ S_\varepsilon ^-=\left\{ {\lambda :\left| \lambda  \right|\le \left( {1+\varepsilon } \right)} \right\}$$
\end{equation}

We use a slightly different contour, a pair of circles around 0 and $\infty$,
and ignore the anti-commuting variables which are irrelevant for understanding
the analyticity of the map.  
We can now factor an analytic map $g$ from $N=1$ superspace into a (super) Lie
group
$G$ into two maps, one of which is analytic
away from 0 and $\infty$, the other meromorphic in neighbourhoods of 
$\left( {0,\infty } \right)$

Let

\begin{equation}
$$X^k=\left\{ \matrix{e:C^*\to G:e,e^{-1}\ have\ Laurent\ expansions\hfill\cr
  of\ order\ k\ and\ e\left( 1 \right)=I\hfill\cr} \right\}$$
\end{equation}

and

$$Y=\left\{ {f:S^2\to G\ meromorphic\ in\ neighbourhoods\ of\ \left( {0,\infty } \right)\ and\ f\left( 1 \right)=I\ } \right\}$$

Any suitable map now has two factorisations,

$$g=f_L^Xf_R^Y=f_L^Yf_R^X$$

where $f_L^X,f_R^X\in X^k$ and $f_L^Y,f_R^Y\in Y$

We use this factorisation to write down a group transformation on the
space $X^k$.

We write $f^\# \left( e \right)=R\cdot e\cdot f$ for $R,f\in Y$ 
and $e,f^\# \left( e \right)\in X^k$.

{\bf Lemma}
If $f^\# \left( e \right)$ can be defined, there exits a unique
$f^\# \left( e \right)$ taking the value $I$ at 1.  Also, if $f^\# \left( 
e \right)$
and $g^\# \left( {f^\# \left( e \right)} \right)$ are defined and normalised to be $I$
at $\lambda = 1$, then

$$\left( {gf} \right)^\# \left( e \right)=g^\# \left( {f^\# \left( e \right)} \right)$$

The proof follows as in Lemma 5.1 of \cite{U}

{\bf Definition} $f\in Y$ is {\bf of simplest type} if 
$f\left( \lambda  \right)=\Pi +\xi \left( \lambda  \right)\Pi ^\bot$,
where $\Pi$ is super-Hermitian projection on complex superspace, 
$\Pi ^\bot = 1-\Pi$ is a projection onto the orthogonal subspace and 
$\xi \left( \lambda  \right)$ is a rational complex function of degree one which is
1 at $\lambda = 1$

{\bf Theorem} 
If $f\left( \lambda  \right)=\Pi +\xi _\alpha \left( \lambda  \right)\Pi ^\bot$
is of simplest type, then $e^\# = f^\# \left( e \right) = R_fef$ is always defined.

To cancel with $f$, $R_f$ must be a similar sum of orthogonal components to cancel
with the corresponding terms:

$$R=\tilde \Pi +\xi \left( \lambda  \right)^{-1}\tilde \Pi ^\bot $$

We need only check that the supersymmetric reality condition
consistently satisfies the requirements of this condition at the zero and pole of $\xi$
which by the reality condition, are given by $\alpha$ and $\lambda ={1 \mathord{\left/ {\vphantom {1 {\bar \alpha }}} \right. \kern-\nulldelimiterspace} {\bar \alpha }}$
respectively.

At $\lambda = \alpha$, we should have

$$\Pi e\left( \alpha  \right)\tilde \Pi ^\bot=0$$

and at $\lambda ={1 \mathord{\left/ {\vphantom {1 {\bar \alpha }}} \right. \kern-\nulldelimiterspace} {\bar \alpha }}$
we need 

\ 
\ 

$$\Pi ^{\bot}e\left( {1 \mathord{\left/ {\vphantom {1 {\bar \alpha }}} \right. 
\kern-\nulldelimiterspace} {\bar \alpha }} \right)\tilde \Pi =0$$

These two equations are compatible, since:
\begin{equation}
$$\matrix{\ \ \left( {\Pi ^\bot e\left( {1 \mathord{\left/ {\vphantom {1 {\bar
\alpha }}} \right. \kern-\nulldelimiterspace}
{\bar \alpha }} \right)\tilde \Pi }
\right)^\# \hfill\cr
  =J^{-1}\left( {\Pi ^\bot e\left( {1 \mathord{\left/ {\vphantom {1 {\bar
\alpha }}} \right. \kern-\nulldelimiterspace}
{\bar \alpha }} \right)\tilde \Pi } \right)^*J\hfill\cr
  =J^{-1}\tilde \Pi ^*e\left( {1 \mathord{\left/ {\vphantom {1 {\bar
\alpha }}} \right. \kern-\nulldelimiterspace}
{\bar \alpha }} \right)^*\Pi ^{\bot *}J\hfill\cr
  =\tilde \Pi J^{-1}e\left( {1 \mathord{\left/ {\vphantom {1 {\bar
\alpha }}} \right. \kern-\nulldelimiterspace}
{\bar \alpha }} \right)^*J\Pi ^\bot \hfill\cr
  =\tilde \Pi \left( {e\left( \alpha  \right)^{-1}} \right)\Pi ^\bot \hfill\cr
  =0\hfill\cr}$$
\label{bign}
\end{equation}

We can now consistently define $\tilde \Pi$ to be the projection on the subspace 
$e\left( \alpha \right)^* V$, where $V$ is the subspace $\Pi$ projects onto.

\subsection{A representation on extended superharmonic maps}

We now have proved that $f\to f^\# $ is a representation on $X^k$.  We can now use this to
define a representation on 
the moduli space of extended solutions of superharmonic map equations 
$E:C^*\times \Omega \to G$. We define
$\left( {f^\# (E)} \right)(q)=f^\# \left( {E(q)} \right)$ for all $q \in \Omega$
and the only condition we need to check is the superharmonic condition:

$$\matrix{\left( {1-\lambda } \right)^{-1}\bar DE\left( {f^\# E}
\right)^{^{-1}}=\tilde A_{\bar \theta }\hfill\cr
  \left( {1-\lambda ^{-1}} \right)^{-1}DE\left( {f^\# E} \right)^{^{-1}}=\tilde
A_\theta \hfill\cr}$$

for some $\tilde A_\theta$, $\tilde A_{\bar \theta}$ independent of $\lambda$.

We use Liouville's theorem to show that the quantities on the left are independent
of $\lambda$ as required. We look at the $\bar D$ term only as the other follows.
This quantity has no pole at 1 because $\left(
{f^\# E} \right)_1\equiv 1$ from before, so that
$D\left( {f^\# E} \right)_1=0$.  Now we can
calculate that

\begin{equation}
$$\matrix{\ \ \ \ \left( {1-\lambda } \right)^{-1}\bar DE\left( {f^\# E} \right)^{^{-1}}\hfill\cr
  =\left( {1-\lambda } \right)^{-1}\bar D\left( {REf} \right)\left( {REf} \right)^{-1}\hfill\cr
  =\left( {1-\lambda } \right)^{-1}\left( {(\bar DR)E+\hat R\bar DE} \right)\left( {RE} \right)^{-1}\hfill\cr
  ={{(\bar DR)R^{-1}} \over {\left( {1-\lambda } \right)}}+\hat R{{\left( {\bar DE} \right)E^{-1}} \over {\left( {1-\lambda } \right)}}R\hfill\cr
  ={{(\bar DR)R^{-1}} \over {\left( {1-\lambda } \right)}}+\hat RA_{\bar \theta }R\hfill\cr
\ \ \ \cr}$$
\end{equation}

The second term is independent of $\lambda$ by Theorem 5.3 because it contains the
superharmonic map term; the first term is holomorphic at $\lambda=0,\infty$ by
construction, so by Liouville's theorem the entire expression is constant in
$\lambda$.  

Again by Theorem 5.3 and the equivalent result for the $D$
term, it follows that $(f^\# E)_{-1}$ is superharmonic.  We must check the normalisation condition,
but since $E(p)\equiv I$, it follows that $R(p)=f^{-1}$ so that $f^\# E = fIf^{-1} =I $, as required.

\subsection{Example Backlund transformations}

We can produce new solutions from known solutions by looking
at just those of simplest
type.

From above, using our representation on extended superharmonic solutions, 
we write for $f=\Pi +\xi \left( \lambda  \right)\Pi ^\bot $,
\begin{equation}
$$f^\# \left( E_\lambda \right)=\left( {\Pi +\xi \Pi ^\bot } \right)E_\lambda \left( {\tilde \Pi +\xi ^{-1}\tilde \Pi ^\bot } \right)$$
\end{equation}

Here $\tilde \Pi $ is the super-Hermitian projection on the subspace 
$E_\alpha ^*V$, where $\alpha$ is the zero of $\xi$ and $V$ is the vector space image of $\Pi$

The new solution is obtained algebraically from $E_\lambda$ but in general this is not trivial.
We can however, find a special case in which the algebra can be calculated to give an easy
relationship between the solutions -- that is the case when our super-uniton solutions have no body, i.e.,
consist of anti-commuting variables only.

To
simplify the calculations, we can use the relationship between the Hermitian
projections defined in (\ref{bign}),
$$\tilde{\Pi}_w E_\alpha^{-1} \Pi_v ^\bot =0$$

where $\tilde {\Pi}_w$ is the Hermitian projection of the transformed solution
given by $\Pi_v$.

From the previous section, we know that in general,
\begin{equation}
$$\Pi _v=\left( {\matrix{{p_v-i\bar vv^\bot \theta \bar \theta }&{i\bar v\bar \theta }\cr
{v^\bot \theta }&{I-iv^\bot \bar v\theta \bar \theta }\cr
}} \right)$$
\end{equation}
where $p_v$ is a holomorphic projector into $U(M)$, and $v$ is the ancillary data
described in the previous section.  Because we are looking only at
the supersymmetric data,
$p_v$ and
$p_w$ are zero.

The extended superuniton solution takes the form
\begin{equation}
$$E_\lambda =\left( {\matrix{{p_u+\lambda \left( {1-p_u} \right)+O(\theta \bar \theta )}&{\left( {1-\lambda } \right)i\bar u\theta }\cr
{\left( {1-\lambda } \right)u^\bot \bar \theta }&{I+O\left( {\theta \bar \theta } \right)}\cr
}} \right)$$
\end{equation}

 In this case the terms in $\theta \bar{\theta}$ have no
effect on the calculation as the anticommuting variables are nilpotent.

Setting $\tilde{\Pi}_wE_\alpha^{-1} \Pi _v^\bot =0$ gives us some algebra to
calculate, but in the end we get a consistent expression relating the two solutions:
\begin{equation}
$$v^\bot =w^\bot \cdot \left( {p+\alpha ^{-1}p^\bot } \right)+u^\bot \left( {1-\alpha ^{-1}} \right)$$
\end{equation}

Using (6.25), and multiplying on the right 
by $\left( {p+\alpha p^\bot } \right)$, 
 we can write down a direct expression for the Backlund transformation, which is
just the inverse: 
\begin{equation}
$$w^\bot =v^\bot \cdot \left( {p+\alpha ^{-1}p^\bot } \right)+u^\bot \left( {1-\alpha ^{-1}} \right)$$
\end{equation}

We thus have an expression for $\Pi_w$ which we can use to write 
down another superharmonic map.   We have therefore proved the following theorem:

{\bf Theorem}  When applied to superunitons, those
maps $f$ of simplest type defined only on superspace produce Backlund
transformations which transform between
 superharmonic maps without body components.

\section{Conclusion and Acknowlegements}

In this paper, we have seen how the harmonic maps into Lie
groups are fundamental theories in several respects, and we have
developed a supersymmetric generalisation.

Using variables and derivates with anti-commuting components, we
have written a super Lax pair and have found special unique
solutions of the supersymmetric chiral model.

The goal in this project is to generalise results obtained in the
classical chiral model theory to the supersymmetric case. The
next step in this effort might be to examine how this theory relates to 
 other known supersymmetric integrable systems, in particular the
supersymmetric Toda models.

A physical analysis of the theory requires us as some point to
Wick rotate the model from the Euclidean case to the Minkowski
metric.  While this raises global difficulties, it should still be
possible to obtain meaningful results locally in the rotated model.

In particular, the Backlund transformations examined in the last
section provide a way to transform between solutions.
In addition, the explicit relationship to Toda theories demonstrated in
\ref{FOD}
should allows us
to calculate physically relevant quantities from the work in the area
of Toda field theory.

Finally, the classification theorem presented here show that the
finite-energy action in the supersymmetric case is discrete
and quantized according to the superuniton number.  The physical
significance of this is yet to be teased out, but the tools
provided here should assist any field-theoretic analysis.

I would like to thank Professor Karen Uhlenbeck for
her patient and skilled supervision over the past number of
years, and to Professor Dan Freed and the rest of my committee
for their interest and support.   I am also indebted to
the University of Texas and the Physics Department for their
enduring practical assistance.

\end{document}